# Optimizing Password Cracking for Digital Investigations

*Mohamad Hachem*
*University of Plymouth, UK*
[mohammadhashem.byte@gmail.com](mailto:mohammadhashem.byte@gmail.com)

*Adam Lanfranchi*
*University of Plymouth, UK*
[adam.lanfranchi@gmail.com](mailto:adam.lanfranchi@gmail.com)

*Nathan Clarke*
*University of Plymouth, UK*
[n.clarke@plymouth.ac.uk](mailto:n.clarke@plymouth.ac.uk)

*Joakim Kävrestad*
Jönköping University
[joakim.kavrestad@ju.se](mailto:joakim.kavrestad@ju.se)

## Abstract

Efficient password cracking is a critical aspect of digital forensics, enabling investigators to decrypt protected content during criminal investigations. Traditional password cracking methods, including brute-force, dictionary, and rule-based attacks, face challenges in balancing efficiency with increasing computational complexity. This study explores rule-based optimization strategies to enhance the effectiveness of password cracking while minimizing resource consumption. By analyzing publicly available password datasets, we propose an optimized rule set that reduces computational iterations by approximately 40%, significantly improving the speed of password recovery. Additionally, the impact of national password recommendations were examined —specifically, the UK National Cyber Security Centre's (NCSC) three-word password guideline—on password security and forensic recovery. Through user-generated password surveys, we evaluate the crackability of three-word passwords using dictionaries of varying common-word proportions. Results indicate that while three-word passwords provide improved memorability and usability, they remain vulnerable when common word combinations are used, with up to 77.5% of passwords cracked using a 30% common-word dictionary subset. The study underscores the importance of dynamic password cracking strategies that account for evolving user behaviors and policy-driven password structures. Findings contribute to both forensic efficiency and cybersecurity awareness, highlighting the dual impact of password policies on security and investigative capabilities. Future work will focus on refining rule-based cracking techniques and expanding research on password composition trends.

## Introduction

Digital forensics is the process of analyzing electronic evidence within criminal investigations and incidents and is crucial to modern crime solving. The role of digital evidence is obvious in cyber-dependent crime such as denial of service attacks or hacking. Digital evidence is equally important in cyber-enabled crime such as drugs and sexual exploitation where evidence such as pictures and chat histories often play a crucial role. However, even in crimes without any obvious digital component, digital evidence can provide important evidence by, for instance, the position where a cellphone has been or investigating a suspect's digital activity before and after a crime (Tart et al., 2021). A digital forensic investigation can be roughly divided into the phases of acquisition, analysis, and reporting (Kävrestad et al. 2024). The acquisition phase is concerned with collecting data and preparing them for forensic analysis. Whether the encrypted data consists of individual files, databases, complete file systems, or entire disks, it is imperative that encrypted content is decrypted in a timely fashion. With widespread use of encryption commonplace, the cracking of passwords required to enable decryption has become a significant barrier.

Password cracking can be broadly classified into brute-force attacks, dictionary and rule-based attacks. A brute-force attack attempts to find a correct password by testing all possible character combinations and is effective against short passwords. The time-complexity of a brute-force attack is exponential with the password length, and it is therefore useless against longer passwords. Without prior intelligence regarding the password composition, brute-force attacks are often not a viable solution for timely cracking. Dictionary attacks are a well-used alternative. A dictionary attack involves building a dictionary of password candidates and then testing them one by one. If the correct password is in the dictionary, it will be cracked. Thus, the effectiveness of dictionary-attacks is dependent on the ability to develop those dictionaries. In practice, forensic experts typically use rule-based attacks to efficiently crack passwords. Rule based attacks are realizations of hybrid attacks where the idea is to broaden a dictionary attack by applying rules to it so that characters are added to dictionary entries, or dictionary entries are modified. A simple example could be to append an explanation mark to all dictionary entries. The intuition behind those attacks is to guess the password behavior of the attack target and develop an attack to mimic that behavior. For instance, if the attack target is likely to use personal details in their password, so should the attack. A successful outcome, the cracking of a password, is therefore dependent on understanding user's password behavior. In practice, a forensic expert would deploy a set of rules, for instance those included in the well-known password cracking tool Password Recovery ToolKit (PRTK) (Exterro, 2025).

The time needed to crack a password is dependent on how many attempts you can make in a reasonable amount of time and investigations are often time sensitive which requires forensic password cracking to be time efficient. In simple terms, the number of passwords that can be tested is a function of the available computational power and the time needed for a single attempt. The computing power available is increasing with the technological advances and one could imagine that would lead to increasing password cracking abilities. However, developers of security functions are increasingly employing techniques to slow down the password cracking process to increase security and it is common that password cracking times are now longer than before, despite more computing power being available. Subsequently, skills in efficient password cracking are more important than even within digital forensics.

A core part of configuring a good rule order is to understand user behavior. It is well-known that people employ different strategies for creating passwords they can remember and use. It is also well-known that those strategies are dependent on how people have been instructed to create strong passwords and are organizationally and regionally dependent. For instance, the UK National Cyber Security Centre (NCSC) recommends composing passwords with three dictionary words. which is likely to impact the password behavior of users in the UK. How this effect impacts forensic password cracking is, however, unknown. The total size of English language in common use is reported to be 170,000 words (English Oxford Dictionary, 2025), which when used in a sequence of three words, theoretically provides for a strong entropy. The overall dictionary is reported to be in excess of 600,000 words. However, individual daily use of the English vocabulary is likely to far smaller. Thereby potentially making cracking simpler.

This paper presents a study investigating password cracking optimization. The research explores two specific research questions:

- Can password cracking rule-sets can be optimized in order to aid timely cracking?
- What is the impact of the NCSC recommendation on using three-word passwords on password cracking?

The paper will begin with an overview of the problem and an analysis of the prior literature. Section 3 presents the research methodology underpinning the research questions and describes the nature of the datasets utilized. The results are presented in section 4. A discussion and conclusion section follow.

## Background Literature

Password security continues to be a cornerstone of modern cybersecurity, with the evolution of password-cracking techniques posing ongoing challenges for security professionals. Over the years, methods have progressed from brute-force and dictionary attacks to more sophisticated approaches such as Markov models, Probabilistic Context-Free Grammars (PCFGs), and neural networks. Despite these advancements, there are still significant challenges in efficiently cracking user- generated passwords, particularly those shaped by complex, mandatory password policies.

The concept of trying all possible combinations of characters to break into a system dates back to the early days of computing. The brute force attack stands out as one of the earliest methods for cracking passwords. Despite being considered outdated against human generated passwords, when it comes to machine-generated or randomly created passwords, no cracking strategy would have an advantage over an exhaustive brute-force attack, where all combinations of a given alphabet, including digits and special characters, up to a predetermined length are tested (Kanta, 2023)

Also considered one of the earliest password-cracking techniques, dictionary attacks involve systematically trying a list of commonly used words or passwords. While straightforward dictionary attacks are limited to exact matches, this technique is still surprisingly successful, as many users tend to choose simple and predictable passwords. These dictionary lists usually include common first and surnames, sports teams, band names, car manufacturers, and famous cities. (Bosnjak et al., 2018). Dictionaries can also be composed of known password lists, with the expectation of password reuse aiding in the successful cracking.

Introduced by Philippe Oechslin in 2003, rainbow tables leverage the concept of time- memory trade-off, using precomputed hash values to crack passwords by significantly reducing the computational cost during the attack. Although this method was groundbreaking at the time, it has become less effective with the advent of salting. Salting adds random data to each password before hashing, ensuring that even if two users have the same password, their hashes will be unique due to the added salt. This technique effectively nullifies rainbow table attacks by creating an infinite number of hash combinations, rendering precomputed tables arguably useless.

However, the evolution of password protection did not stop with salting alone. To further slowdown brute-force and dictionary attacks, modern systems incorporate Key Derivation Functions (KDFs), which strengthen password hashing mechanisms by adding computational complexity. Key Derivation Functions (KDFs) are cryptographic functions designed to convert input values like passwords into secure cryptographic keys used for data encryption. A KDF typically takes a password, salt (random data), and configuration parameters such as the number of iterations, and generates a key that can be used to protect sensitive data. One of the main purposes of a KDF is to add computational complexity to the process of generating keys from passwords, making it harder for attackers to reverse-engineer or guess the original password by brute force. Commonly used KDFs, such as PBKDF2 (Password-Based Key Derivation Function 2) and Argon2, are integral to enhancing security in password-based systems by ensuring that even if the password itself is weak, the derived key remains more resistant to attacks due to the KDF's design (Kaliski, 2000; Biryukov et al., 2016).

KDFs strengthen the security of password hashes by making brute-force and dictionary attacks more computationally expensive. Hashing alone is not enough to secure passwords, as fast hashing algorithms can allow attackers to attempt millions of guesses per second using modern hardware like GPUs. KDFs introduce intentional computational delays by performing multiple iterations of a hash function, ensuring that each attempt to guess a password consumes significantly more resources. For instance, PBKDF2 performs a large number of iterations of the hash function, which makes each password guess take longer to compute (Chen, 2024). Argon2, a newer KDF, not only increases computational complexity but also memory usage, making it difficult for attackers to exploit parallel computing resources effectively. (Biryukov et al., 2016)

By increasing the time and resources required to compute each password guess, KDFs drastically slow down the speed at which password crackers can test potential passwords. This reduces the effectiveness of password cracking approaches – often resulting in a brute-force technique being not viable. The added complexity means that attackers must invest much more time and computing power to break a password, significantly lowering the chances of success. In contrast to simple hash functions, KDFs like PBKDF2, Argon2, and bcrypt ensure that even common passwords are more difficult to crack due to their computationally intensive nature.

# Methodology

This research used a mixed-method approach where an experimental research design was used to evaluate the optimal order of rules for rule-based password cracking. A subsequent survey-based approach sought to investigate user's password composition practices, including the use of the three-word password recommendation by NCSC on password composition among British users to study the impact of regional recommendations on password composition.

### *Optimizing the order of password cracking rules*

To analyze the optimal order of rules for password cracking we used an experiment where we evaluated the effectiveness of each individual rule in the password-cracking tool Password Recovery ToolKit (PRTK). PRTK is a well-established commercial tool provided as part of the Forensic Toolkit suite (Exterro, 2025). Using several datasets that have been released from prior attacks, it was possible to identify the rules that were more likely to be successful over others, whilst at the same time measuring the computational requirements of the rule. Using this information, the order of the rules was modified in the most optimal way (with the most effective rule on top) and validated using unseen data(sets). The results where then compared with the PRTK default.

To help evaluate whether the optimization would work effectively, the study used three publicly available password lists, based on suggestions from the existing literature (Jin and Dupuis 2024; Lee, et al., 2022; Piotr, 2021):

- piotrcki-workdlist-top10m which contains close to 10,000,000 passwords from haveibeenpawned
- Rockyou which contains 14,341,564 passwords and is commonly used for password cracking and research
- xato-net-1million-passwords – published by the Xato network who specialize in creating datasets for cybersecurity research.

As illustrated in Table 1, two of the datasets were split in order to provide a mechanism of optimizing the order of the rules, with the unseen second parts of each dataset being used to evaluate the optimization. A third dataset, not used in configuring the optimisation was also utilized to examine how well the approach would work against a completely unseen dataset.

**Table 1: Password Dataset Configuration**

| Dataset | Subset | Purpose | Sample Size |
|---|---|---|---|
| Piotrcki_Wordlist_top10m | Pio1 | Optimization | 5,000,000 |
|  | Pio2 | Evaluation | 5,000,000 |
| Rockyou | Rock1 | Optimization | 7,170,782 |
|  | Rock2 | Evaluation | 7,170,782 |
| Xato_net_1m_passwords | Xato | Evaluation | 1,000,000 |

### *Impact of national password recommendations*

To analyze the potential impact of the NCSC recommendation of using passwords composed of three-words on password cracking, it was necessary to collect data – as no available dataset with this password composition could be found. A survey was developed and distributed to participants using a multitude of channels such as direct email, social media and online communities. While this creates a convenience sample, it was deemed a good approach for maximizing the number of respondents (Henry 1990). The survey contains three parts; Demographic information, self-reported password behavior, and one section where the participants were asked to create passwords follows different sets of guidelines. The study then assessed the strength of the created passwords and compared them to the self-reported password behavior.

The survey collected data about the participants background and password behavior and asked the participants to create passwords given different password advice such as traditional alphanumerical passwords, and three random words password as per the suggestion from NCSC. Four passwords were created by each participant with the following password requirements:

- At least eight-character long passwords containing upper- and lower-case characters
- At least eight-character long passwords containing upper- and lower-case characters and numbers
- At least eight-character long passwords containing upper- and lower-case characters, numbers and special characters
- Three dictionary words

A final single password with the following composition was also required:

- Three dictionary words of at least eight characters each – where we forced the length on the dictionary word in order to understand and compare the impact of doing so.

The survey was distributed via several online channels and a total of 231 individuals participated, creating a total of 3,927 passwords. The results were analyzed with a focus on the following aspects:

- Examining the strength of three-word passwords and how this compares to normal composition approaches
- Per-user comparison of the strength of random and three-word passwords

For establishing the strength of password compositions, participant entries were compared to the Rockyou2024 dataset, comprising of 10 billion entries (Price, 2024). This would aid in establishing whether the password could be cracked with ease. Given the size of the English language dictionary and the use of three-dictionary words, it is not computationally feasible to brute force the space in order to determine the effort required to crack. Instead, the study developed three dictionary datasets, one based upon 10% of the most common English language words, 20% and 30%. These dictionaries were felt to be of an appropriate size to permit timely cracking in practice. These dictionaries were identified and downloaded from the British National Corpus (BNC). The Brown corpus was used (Natcorp, 2024).

## Results

This section will outline the results from the two experiments.

### *Optimizing the order of password cracking rules*

The first experiment consisted of two stages: evaluation of individual rule and composition and validation of an ideal rule-order. The datasets were prepared by splitting the datasets piotrcki-workdlist-top10m and Rockyou into two equally large datasets. Those were denoted PIO1 and PIO2, and Rock1 and Rock2. PIO1 and Rock1 were used in the first stage to model the effectiveness of each rule, while the other datasets were used in the second stage. Table 2 presents the default rule-order provided by PRTK.

In the first experiment, we evaluated how many of the passwords in the datasets PIO1 and ROCK1 that matched the respective rules included in PRTK default attacks, the rules included in the attack and their respective order is shown in Table 3. This was achieved by using a script that evaluated the passwords to see if they matched any of the rules. Note that rules which matched less than 0.5% of the passwords in any of the datasets are omitted to save space.

**Table 2 PRTK default rule order**

| Position | Pattern | Position | Pattern | Position | Pattern | Position | Pattern |
|---|---|---|---|---|---|---|---|
| 1 | One-digit | 2 | One_letter | 3 | Two_digits | 4 | Two_letters |
| 5 | Three_digits | 6 | Three_letters | 7 | Four_digits | 8 | Five_digits |
| 9 | Six_digits | 10 | Four_letters | 11 | Five_letters | 12 | Seven_digits |
| 13 | eight_digits | 14 | Two_uppercase | 15 | Digit_between_letters | 16 | All_lowercase |
| 17 | One_character | 18 | Two_characters | 19 | Three_characters | 20 | Four_characters |
| 21 | Digit_4letters | 22 | 4letters_digit | 23 | 2letters_3digits | 24 | 2digits_3letters |
| 25 | 3letters_2digits | 26 | 4letters_symbol | 27 | Symbol_4letters | 28 | Six_letters |
| 29 | 2digits_4letters | 30 | 2letters_4digits | 31 | 3letters_3digits | 32 | 4letters_2digits |
| 33 | 2digits_letters | 34 | Letters_2digits | 35 | Letters_4digits | 36 | 4digits_letters |
| 37 | Any_5characters | 38 | 4letters_3digits | 39 | any_6characters | 40 | Seven_letters |
| 41 | any_7characters | 42 | Eight_letters | 43 | any_8characters | 44 | Nine_letters |
| 45 | any_9characters | 46 | Ten_letters | 47 | Any_10characters | 48 | any_11characters |
| 49 | Any_12characters | | | | | | |

**Table 3 Passwords matched per PRTK rule**

| Rule | Percentage in PIO1 | Percentage in Rock1 |
|---|---|---|
| All_lowercase | 19% | 25.2% |
| Letters_2digits | 16.3% | 13.7% |
| 4letters_2digits | 2.4% | 1.7% |
| Letters_4digits | 7.4% | 7.9% |
| 2letters_4digits | 0.8% | 0.7% |
| Six_digits | 6.8% | 2.7% |
| Two_uppercase | 3% | 4.6% |
| Seven_digits | 2% | 3.3% |
| Eight_digits | 1.8% | 2.9% |
| 4letters_3digits | 1% | 0.8% |
| 3letters_3digits | 0.9% | 0.7% |
| Any_10characters | 8.6% | 7.4% |
| Any_9characters | 7.4% | 7.2% |
| Any_8characters | 7.4% | 6.3% |
| Any_7characters | 4.6% | 4.3% |
| Any_6characters | 2.6% | 2.3% |

Based on the results, a custom rule order was developed with the intent of using the rules with the highest chance of success first. The rationale is that the rules are applied in order so that a password matching the last rule will first need to go through all the previous rules before its cracked. The custom rule order is displayed in Table 4.

**Table 4 Custom rule order**

| Position | Pattern | Position | Pattern | Position | Pattern | Position | Pattern |
|---|---|---|---|---|---|---|---|
| 1 | All_lowercase | 2 | 4letters_2digits | 3 | 2letters_4digits | 4 | 3letters_2digits |
| 5 | Letters_2digits | 6 | Letters_4digits | 7 | Six_digit | 8 | Two_uppercase |
| 9 | Seven_digits | 10 | Eight_digits | 11 | 4letters_3digits | 12 | Five_letters |
| 13 | Five_digits | 14 | 3letters_3digits | 15 | 4digits_letters | 16 | 2digits_4letters |
| 17 | 2digits_3letters | 18 | 2digits_letters | 19 | Four_letters | 20 | Six_letters |
| 21 | Three_letters | 22 | Digit_between_letters | 23 | Seven_letters | 24 | Eight_letters |
| 25 | Nine_letters | 26 | 4letters_digit | 27 | Four_digits | 28 | Ten_letters |
| 29 | Two_letters | 30 | 2letters_3digits | 31 | Three_digits | 32 | Digit_4letters |
| 33 | 4letters_symbol | 24 | Symbol_4letters | 25 | Two_digits | 36 | One_letter |
| 37 | One_digit | 28 | Any_10characters | 39 | Any_9characters | 40 | Any_8characters |
| 41 | Any_7characters | 42 | Any_6characters | 43 | Any_11characters | 44 | Any_12characters |
| 45 | Three_characters | 46 | Four_characters | 47 | Any_5characters | 48 | Two_characters |
| 49 | One_character | | | | | | |

The custom rule order was evaluated using the second sections of the Pio and Rockyou datsts as well as the completely unused Xato dataset. As seen in Table 5, The results for the PRTK default rule set and the custom rule set is identical, which is unspringing since the same rules are used but does help to provide reassurance on the integrity of the methodology and scripts.

**Table 5 Passwords founds per rule set**

| Dataset | Total Passwords Found (Custom) | Percentage (Custom) | Total Passwords Found (PRTK) | Percentage (PRTK) |
|---|---|---|---|---|
| Pio2 | 4,897,054 | 97.94% | 4,897,054 | 97.94% |
| Rock2 | 6,919,345 | 97.36% | 6,919,345 | 97.36% |
| Xato | 987,763 | 98.78% | 987,763 | 98.78% |

The hypothesized benefit of the custom rule order is that it can improve efficiency. This hypothesis was confirmed by analyzing the number of iterations needed by each rule order. The number of iterations is the number of password guesses needed to find all passwords. It is used in this experiment as a measure of efficiency of the rule set. The number of iterations needed for each rule set is shown in Table 3.

**Table 6 Iterations consumed per rule set**

| Rule Order | Iteration Count | | |
| --- | --- | --- | --- |
| | PIO2 | Rock2 | Xato |
| PRTK | 149,508,390 | 213,513,180 | 24,911,595 |
| Custom | 85,798,632 | 126,498,465 | 14,027,532 |
| Efficiency of Custom | 42.63% | 40.71 % | 41.76% |

As shown in Table 3, our custom rule order consumes about 40% less iterations than the PRTK default for all three data sets, which will result in more timely and efficient cracking. Table 7 provides a further analysis of how the two rulesets crack passwords, set at applying the first 2, 5, 10 and 15 rules. The results not only show that ordering the rules provides for a high proportion of password cracking, but this can also be achieved at a significantly reduced iteration count. For example, the custom order has cracked 72% of passwords in the Xato dataset with only 7.25 million iterations, versus the default order only achieving 25% but with 14 million iterations.

**Table 7 Passwords and iterations for the Xato Dataset against number of rules**

| Rule Order | # of Rules | Passwords Matched | % of Total | Iteration Count |
| --- | --- | --- | --- | --- |
| PRTK | 2 | 5 | 0.0005 | 2,000,000 |
| **Custom** | **2** | **353,697** | **35.4** | **1,600,000** |
| PRTK | 5 | 232 | 0.023 | 5,000,000 |
| **Custom** | **5** | **454,000** | **45.4** | **3,500,00** |
| PRTK | 10 | 103,080 | 10.3 | 10,000,000 |
| **Custom** | **10** | **682,619** | **68.3** | **5,700,000** |
| PRTK | 15 | 245,000 | 24.5 | 14,000,000 |
| **Custom** | **15** | **717,000** | **71.7** | **7,250,000** |

*Impact of national password recommendations*

To establish the relative strength of traditionally composed passwords versus the three-word rule, the captured passwords were hashed and then submitted to CrackStation (2025) and Hashes.com (2025) -two online hash cracking tools to establish passwords could be cracked with relative ease. Table 8 presents the results set against each of the password composition rules. The results show that the three-word rule remains relatively robust against the traditional rules, sitting between the eight character, upper/lower case, numbers with and without a special character. When participants were forced into using dictionary words of eight characters or more, cracking reduced to 3.3% and was the strongest of the rules provided.

**Table 8 Proportions of cracked passwords per password style**

| Password type | % cracked |
| --- | --- |
| Eight-character long passwords containing upper- and lower-case characters | 32.6% |
| Eight-character long passwords containing upper- and lower-case characters and numbers | 16.0% |
| Eight-character long passwords containing upper- and lower-case characters, numbers and special characters | 3.9% |
| Three words | 11.4% |
| Three eight-character words | 3.3% |

The three word and three eight-character word datasets were also compared against the 10, 20, 30% most common English dictionary sets to identify the ease to which they could be cracked. Table 9 presents the results of the analysis, which show a signification proportion of the passwords can be established with a relatively small proportion of the English dictionary. Interestingly, whilst encouraging users to use longer eight-character or more dictionary words led to a higher proportion of password cracking when using the 20% and 30% dictionary sample.

**Table 9 Cumulative percentage of three word passwords cracked**

| Password type | 10% most common English words | | 20% most common English words | | 30% most common English words | |
|---|---|---|---|---|---|---|
| | % Found | Time (mins) | % Found | Time (mins) | % Found | Time (mins) |
| Three words | 17.5 | 132 | 38.6 | 1533 | 50.9 | 4999 |
| Three eight-character words | 10.4 | 125 | 44.2 | 1501 | 77.5 | 5053 |

Analyzing the time taken to brute force the respect dictionaries, it is clear there is a significant increase in computation moving from 10% and a worse case time of 132 minutes to exhaust, to 30% and a worse case time of 5053 minutes to compute. As the proportion of most common words increases, it would be expected to see a similar significant increase in the time required. Whilst this helps to support the rationale for using such a rule, with those passwords that were not cracked potentially being very challenging to recover, it does not remove the fact that half to three quarters of passwords can be recovered utilizing these smaller common-word English dictionaries.

## Discussion

The results have highlighted both the challenges with effective password cracking and opportunities that are provided when examining and optimizing the password cracking rules. Table 10 provides an analysis of the time that would be saved based upon the optimization of rules. This was calculated based upon on average of 80 hash comparisons per second (derived from a real-world test using PRTK on a PC with Intel i7-13700K and 32GB RAM). In investigations when time is critical or computational resource limited, these savings can have a significant impact.

**Table 10 Time saved across the three datasets**

| Rule Order | PIO2: Iteration count | Rock2: Iteration count | Xato: Iteration count |
|---|---|---|---|
| PRTK | 149,508,390 | 213,513,180 | 24,911,595 |
| Custom | 85,798,632 | 126,498,465 | 14,027,532 |
| **Time saved** | **298 hours** | **302 hours** | **37.8 hours** |

It is very likely that password compositions will vary over time, necessitating a revised analysis and update to the optimized rulesets. As shown in the UK-example, specific password policies will also impact the utilization of certain rulesets and it's important for forensic investigators to be mindful of these differences when configuring password cracking tools.

Digital forensic investigators have long benefitted from individual's inability to manage and utilize strong passwords. The usability issues surrounding these are well documented and have directly led to many of the rulesets that are in common use today. The NCSC recommendations are amongst the first globally to recognize the challenges in both providing secure and usable passwords – which in one sense, should be commended. However, the results presented in this study also highlight the serious implications of this policy. For forensic investigators, it clearly presents a new opportunity for successfully cracking passwords. But this opportunity will also be extended to hackers, giving rise to the potential for successful attacks – which is bad news for both individuals and digital forensic investigators, as it will lead to increase incidents that need investigating.

It is worth also reflecting upon some of the limitations of this study, to aid future directions of research:

- It was not possible to implement all of the PRTK rulesets with the experiment. A number of the more complicated Markov-based approaches could not be implemented. The prior literature also points to AI-based approaches as being effective. These can be incorporated into this approach as additional rules and should be evaluated.
- The datasets utilized in this study are reflective of normal users – both the captured passwords from password hacks and in the survey where individuals were asked to compose passwords of varying compositions. Motivated and knowledgeable attackers are likely to use stronger passwords in practice than those from the general population. Some care is therefore required in interpreting how effective password cracking of suspects can be. Law Enforcement Agencies (LEAs) will have a golden dictionary of previously cracked passwords from suspects. If access were possible, profiling and optimizing based upon these should provide a more effective result.
- Due to the computational limitations and time for this study, it was not possible to extend the password cracking of three-dictionary words beyond 30%. Going further and exploring the impact of larger proportions of the dictionary would be relevant and interesting to investigate. This might aid in establishing what the size of the "usable" dictionary is for people, which would also help to provide more reliable estimates of the entropy of the approach in practice.

## Conclusion and Future work

The study sought to explore whether password cracking could be made more efficient through an examination of the rulesets that are applied. Both studies have shown that the success of password cracking and the time required to crack can be significantly improved through optimization – whether that be through rulesets or the use of the top *x* percent of common words. The use of optimized rulesets could save 38-302 hours of computation based upon the datasets utilized – which is a significant amount of time, resource and money.

The challenge is in ensuring the optimization is based upon current practice. Obtaining real-world examples of passwords from relevant users (i.e. suspects as they tend to be less willing to provide the password) will somewhat limit the effectiveness. However, the results have shown a significant step-change in performance that it would be expected that this should follow-through – after all, human behavior will also be exhibited by suspects.

Future research will focus further exploring the role of organization and national policy and culture on password composition strategies. This has the potential to aid in the development of more localized ruleset optimizations that take this into consideration, enabling investigators to develop more efficient and granular password cracking.